\RequirePackage{fix-cm}

\documentclass[smallextended]{svjour3}

\usepackage{url}
\usepackage[authoryear]{natbib}
\usepackage{dirtytalk}
\usepackage{tcolorbox}
\usepackage{booktabs}
\usepackage{multirow}
\usepackage{tabularx}
\usepackage{fontawesome5}
\usepackage{orcidlink}
\usepackage{framed}
\usepackage[dvipsnames]{xcolor}
\usepackage{graphicx}
\usepackage{adjustbox}
\usepackage{makecell}
\usepackage{listings}
\usepackage{array}

\newcolumntype{Y}{>{\centering\arraybackslash}X}

\title{Beyond Strict Rules: Assessing the Effectiveness of Large Language Models for Code Smell Detection}
\titlerunning{Effectiveness of LLMs for Code Smell Detection}
\authorrunning{Souza et al.}

\author{
Saymon Souza\thanks{Corresponding author: silvasouzasaymon@gmail.com} \and
Amanda Santana \and
Eduardo Figueiredo \and
Igor Muzetti Pereira \and
Jo{\~a}o Eduardo Montandon \and
Lionel Briand
}

\institute{
Saymon Souza \at
Universidade Federal de Minas Gerais, Belo Horizonte, Minas Gerais, Brazil \\
\email{silvasouzasaymon@gmail.com}
\and
Amanda Santana \at
Universidade Federal de Minas Gerais, Belo Horizonte, Minas Gerais, Brazil
\and
Eduardo Figueiredo \at
Universidade Federal de Minas Gerais, Belo Horizonte, Minas Gerais, Brazil
\and
Igor Muzetti Pereira \at
Universidade Federal de Ouro Preto, Ouro Preto, Minas Gerais, Brazil
\and
Jo{\~a}o Eduardo Montandon \at
Universidade Federal de Minas Gerais, Belo Horizonte, Brazil
\and
Lionel Briand \at
University of Ottawa, Ontario, Canada.
}

\journalname{Empirical Software Engineering}

\setcitestyle{aysep={}}

\lstdefinestyle{codeStyle}{
  language=Java,
  basicstyle=\ttfamily\small,
  frame=single,
  breaklines=true,
  columns=fullflexible,
  keepspaces=true,
  numbers=left,
  numberstyle=\tiny\color{gray},
  stepnumber=1,
  numbersep=8pt,
  keywordstyle=\color{blue}\bfseries,
  commentstyle=\color{gray}\itshape,
  stringstyle=\color{orange},
  emph={visitFieldInsn,visitMethodInsn},
  emphstyle=\color{teal}\bfseries
}

\begin{document}

\date{Received: date / Accepted: date}

\maketitle

\begin{abstract}
    Code smells are symptoms of potential code quality problems that may affect software maintainability, thus increasing development costs and impacting software reliability. Large language models (LLMs) have shown remarkable capabilities for supporting various software engineering activities, but their use for detecting code smells remains underexplored. However, unlike the rigid rules of static analysis tools, LLMs can support flexible and adaptable detection strategies tailored to the unique properties of code smells. This paper evaluates the effectiveness of four LLMs -- DeepSeek-R1, GPT-5 mini, Llama-3.3, and Qwen2.5-Code -- for detecting nine code smells across 30 Java projects. For the empirical evaluation, we created a ground-truth dataset by asking 76 developers to manually inspect 268 code-smell candidates. Our results indicate that LLMs perform strongly for structurally straightforward smells, such as \textit{Large Class} and \textit{Long Method}. However, we also observed that different LLMs and tools fare better for distinct code smells. We then propose and evaluate a detection strategy that combines LLMs and static analysis tools. The proposed strategy outperforms LLMs and tools in five out of nine code smells in terms of F1-Score. However, it also generates more false positives for complex smells. Therefore, we conclude that the optimal strategy depends on whether Recall or Precision is the main priority for code smell detection.
\end{abstract}

\keywords{Code Smells \and LLM \and Empirical Study \and Tools \and Software Quality}

\section{Introduction}
\label{sec:introduction}

Code smells are symptoms or indicators of potential code quality problems that may affect software maintainability and reliability \citep{Fowler1999}. Code maintainability is essential because it refers to how easily code can be understood, changed, and improved~\citep{sjoberg2012quantifying}. In this context, previous research has shown that code smells can increase development costs~\citep{abbes2011empirical, palomba2014they}, reduce software reliability~\citep{xia2017measuring,yamashita2013code}, and lead to software defects~\citep{hall2014some,khomh2012exploratory,olbrich2010all,palomba2018diffuseness}. Detecting and refactoring code smells can be challenging because manual reviews require advanced skills, are expensive and slow, while automated tools are imprecise and still require human interpretation~\citep{palomba2014they}. Common examples of code smells include methods that become complex because they take on too many responsibilities, intense communication between classes, or the same code snippets repeated in different places~\citep{Fowler1999}.

Recent advances in Large Language Models (LLMs) have sparked interest in their use for coding problems~\citep{al2022readable,mastropaolo2023robustness,obrien2024prompt}. In fact, these models have shown promising capabilities for generating code~\citep{khan2022automatic}, repairing bugs~\citep{choi2024iterative}, and supporting software testing~\citep{wang2024software}, but the challenges of using LLMs to detect code smells remain underexplored and lack relevant benchmarks~\citep{wu2024ismell}. In contrast to traditional automated detection tools that adhere to strict rules, such as JDeodorant~\citep{tsantalis2008jdeodorant} and~\citet{PMD}, LLMs can provide an innovative approach to detecting code smells. Their ability to understand complex contexts may enable them to adopt flexible, adaptable strategies for different types of code smells. In fact, some preliminary studies have investigated the use of LLMs to detect and refactor code smells~\citep{jiang2024unearthing, wu2024ismell}. However, to the best of our knowledge, they do not provide strong empirical evidence to support, for instance, the claim that LLMs perform better than traditional code smell detection tools.

Moreover, early investigations of LLMs for code smell detection typically employ a small set of smell types and simple code samples~\citep{dakhel2023github,liu2024no,tufano2022using}. Thus, it is essential to further investigate the effectiveness and limitations of LLMs for detecting code smells in real-world software projects~\citep{wu2024ismell}. Such software projects introduce numerous challenges for LLMs, as they require navigating large, complex codebases, adhering to diverse coding standards, and ensuring compatibility with existing systems~\citep{nunes2024}. These characteristics make automated detection of code smells particularly difficult, especially when compared to controlled or synthetic code examples. Evaluating how effectively LLMs can detect code smells under these realistic conditions is therefore crucial, as it helps determine whether they can serve as practical alternatives, or valuable complements, to traditional detection tools~\citep{silva2024detecting}. More importantly, we need to empirically determine not only how LLMs compare with other automated techniques, but also how closely their detections align with human judgments of what constitutes a code smell.

This paper evaluates the effectiveness of LLMs by using four models, namely DeepSeek-R1, GPT-5 mini, Llama-3.3, and Qwen2.5-Coder, to detect nine code smells in 30 real-world Java projects. \textbf{Our goal is to understand which code smells LLMs successfully detect and when they fail}. To perform our empirical evaluation, we first expanded a large dataset of 30 top-starred Java projects mined from GitHub~\citep{Santana2024}. Since the previous dataset included only smells detected with static analysis tools, we not only expanded it with new LLM and human evaluations but also created ground truth by asking 76 developers to manually inspect 268 code-smell candidates.

Using the newly expanded dataset as the foundation of our investigation, we conducted three complementary analyses to assess the effectiveness of LLMs in detecting code smells. The main analysis evaluates how accurately each LLM identifies code smells relative to the human-validated ground truth derived from developer evaluations. Building on this foundation, the second analysis compares LLM performance with that of traditional static analysis tools to identify where LLMs offer advantages or show limitations. Finally, the third analysis examines whether combining both detector types using a voting-based strategy can further improve detection accuracy and coverage.

Our results indicate that LLMs perform strongly for structurally simple smells, such as \textit{Large Class} and \textit{Long Method}, especially when multiple detection strategies are combined via voting. This proposed strategy that aggregates outputs of four LLMs and two static analysis tools maximizes recall and F1-score. For more subjective or context-dependent smells, such as \textit{Feature Envy} and \textit{Refused Bequest}, LLMs provide mixed results, although specialized tools or carefully selected individual LLMs remain the most effective. The results also show that no single strategy outperforms the others across all smells. In most cases, the optimal strategy varies depending on the smell type. In particular, the combined strategy consistently improves recall and provides robust detection of common, easily quantifiable smells, making it an attractive choice for practitioners seeking greater code-smell coverage. However, this combined strategy may generate more false positives for complex smells. As a result, the choice between a combined and an individual detection strategy should reflect the project’s tolerance for such trade-offs and its specific code quality goals.

The contributions of this work can be summarized as follows.
\begin{itemize}
  \item We created a ground truth of code smells in 30 real-world Java projects mined from GitHub. This ground truth can be used to compare the effectiveness of different strategies to detect code smells, including other static analysis tools and LLMs beyond those evaluated in our study.
  \item We demonstrate that a combined strategy offers the best recall and F1-score for detecting 5 out of 9 code smells. However, specialized tools or LLMs remain preferable for more subjective cases, underscoring the importance of selecting detection strategies based on the smell type and the desired trade-off between Recall and Precision.
  \item For software developers, we indicate which automated strategy is more effective in detecting each type of code smell. Our results also help teams refine their workflows by balancing Recall, Precision, and review costs.
  \item We provide a dataset and all scripts to replicate and expand this study, for instance, with other artificial intelligence models and tools. The online artifacts used in this study are available in our replication package~\citep{dataset}.
\end{itemize}

The remainder of this paper is organized as follows. Section~\ref{sec:background} introduces code smells and describes the previous dataset, which is extended in our study. Section~\ref{sec:dataset} outlines how we extended the previous dataset and created our ground truth. Section~\ref{sec:research-method} explains our research method, and Section~\ref{sec:results} presents our empirical results. Section~\ref{sec:discussion} analyzes the main findings of this study. Section~\ref{sec:threats} discusses possible threats to validity and the actions we took to mitigate them. 
Section~\ref{sec:related-work} discusses some related work.
Finally, Section~\ref{sec:conclusion} provides our final thoughts and suggests ideas for future research work.

\section{Background}
\label{sec:background}

This section presents an overview of the code smells analyzed and the dataset used in this study.

\subsection{Code Smells and Detection Techniques}
\label{sec:background:smells}

\textit{Code smells} are symptoms or indicators of potential code quality degradation that may affect the software maintainability and reliability~\citep{Fowler1999}. They not only impact code understandability, reusability, and extensibility~\citep{sjoberg2012quantifying,yamashita2013code}, but they may also be the source of bugs and code instability~\citep{hall2014some,palomba2018diffuseness,santos2023yet}. Given their potential to increase development costs~\citep{sjoberg2012quantifying}, it is important to identify code smells and refactor them. \textit{Refactoring} is an activity in which the code is modified to improve its internal quality without changing its external behavior~\citep{Fowler1999}. Table~\ref{tab:smell_definition} describes the nine code smells used in this study. The first column shows the smell name, while the second column briefly defines each code smell. More details on their definition can be found in the books of~\citet{Fowler1999}, and~\citet{lanza2007object}. These particular code smells were chosen because they are well supported by detection tools~\citep{PMD,tsantalis2008jdeodorant} and cover a broad range of modularity-related issues~\citep{fernandes2016review}. 

\begin{table}[ht]
  \centering
  \caption{Code Smell Definitions}
  \label{tab:smell_definition}
  \footnotesize
  \setlength{\tabcolsep}{4pt}   
  \renewcommand{\arraystretch}{1.1}

  \begin{tabularx}{\columnwidth}{@{}l X@{}}
    \toprule
    \textbf{Code Smells} & \textbf{Definitions} \\
    \midrule
    Data Class
      & A class composed mainly of fields and getter/setter methods, with
        little or no meaningful behavior.~\citep{Fowler1999} \\

    Dispersed Coupling
      & A method that depends on many other classes, but with low
        coupling intensity.~\citep{lanza2007object} \\

    Feature Envy
      & A method that accesses members of other classes more than its own.~\citep{Fowler1999} \\

    Intensive Coupling
      & A method that heavily interacts with one or a few other classes,
        forming a tight cluster.~\citep{lanza2007object} \\

    Long Method
      & A method that is excessively long or takes on too many
        responsibilities.~\citep{Fowler1999} \\

    Large Class
      & A class that handles multiple responsibilities or contains 
        many lines of code.~\citep{Fowler1999} \\

    Long Parameter List
      & A method signature that requires an excessive number of
        parameters.~\citep{Fowler1999} \\

    Refused Bequest
      & A subclass that overrides or ignores most inherited behavior,
        indicating poor inheritance fit.~\citep{Fowler1999} \\

    Shotgun Surgery
      & A change in one module forces many small changes scattered
        across other modules.~\citep{Fowler1999} \\
    \bottomrule
  \end{tabularx}
\end{table}

Several techniques to detect code smells have been proposed in the literature, such as manual code inspection~\citep{madeyski2023detecting}, static analysis tools~\citep{PMD,tsantalis2008jdeodorant}, refactoring opportunities~\citep{fokaefs2011jdeodorant}, change history analysis~\citep{palomba2013hist}, and machine learning models~\citep{di2018detecting,madeyski2023detecting,santos2023yet}. Although developers widely use static analysis tools to detect code smells~\citep{tsantalis2008jdeodorant}, several studies in the literature indicate they have poor agreement with developers' perception of what a code smell is~\citep{fernandes2016review, Yamashita2013}. Previous work also indicates the low effectiveness of some classic machine learning models, such as Naive Bayes, Decision Tree, and Random Forest, to detect code smells~\citep{cruz2020detecting, di2018detecting, nunes2024}. More importantly, although some recent preliminary studies have been published on these code smell detection techniques~\citep{jiang2024unearthing,wu2024ismell}, we still lack strong empirical evidence on the effectiveness of LLMs in supporting code smell detection.

\subsection{Datasets of Code Smells}
\label{sec:background:original-dataset}

Several datasets of code smells are available in the literature ~\citep{cruz2020detecting, madeyski2023detecting, Tempero2010}. However, they have several limitations. For instance, systems in some datasets may not reflect current development practices ~\citep{cruz2020detecting,Tempero2010}, and they exhibit limited coverage of code smells ~\citep{madeyski2023detecting}. To avoid these limitations, we have used and extended our previous dataset~\citep{Santana2024}. Our dataset extension includes additional features, the integration of LLMs for code smell detection, and human evaluations (see Section~\ref{sec:dataset}). The selected dataset included 3,459 instances of 9 code smells (3 at the class level and 6 at the method level) across 30 open-source Java systems on GitHub. We focus on Java projects for this study because many tools are available to detect a wide range of code smells in this programming language~\citep{fernandes2016review}.

Table~\ref{tab:final-dataset} provides details about the 30 systems included in the used dataset~\citep{Santana2024}. The first column lists the names and versions of each system. The following three columns show the number of classes (NOC), the number of methods (NOM), and the lines of code (LOC) for each system, respectively. The last column (Stars) displays the total number of stars for each repository. The following criteria were used to select the systems: (i) they were among the top-stared Java systems on GitHub; (ii) they had commits merged in the last two years; and (iii) they cover different domains, sizes, and levels of maturity. Systems for educational purposes were excluded~\citep{Santana2024}. This variety of systems helps us investigate how well LLMs detect code smells across software systems of different domains and sizes. 

\begin{table}[ht]
  \centering
  \caption{Dataset Description}
  \label{tab:final-dataset}
  \begin{tabular}{l r r r r}
    \toprule
    \textbf{Name} & \textbf{NOC} & \textbf{NOM} & \textbf{LOC} & \textbf{Stars} \\
    \midrule
    arthas-3.4.3 & 834 & 4,733 & 39,973 & 36,580 \\
    cryptomator-1.6.1 & 590 & 2,690 & 16,350 & 13,388 \\
    dbeaver-21.0.2 & 6,449 & 36,575 & 348,608 & 44,710 \\
    easyexcel-2.2.11 & 249 & 1,629 & 10,639 & 33,549 \\
    elasticsearch-analysis-ik & 28 & 203 & 2,051 & 17,200 \\
    fastjson-1.2.76 & 249 & 1,996 & 44,434 & 25,756 \\
    gson-2.8.8 & 231 & 924 & 11,721 & 23,925 \\
    guava-30.1.1 & 27,412 & 200,616 & 2,125,859 & 50,993 \\
    HikariCP-4.0.0 & 68 & 581 & 4,530 & 20,634 \\
    hutool-5.7.17 & 1,214 & 11,966 & 79,432 & 29,969 \\
    java-faker-1.0.2 & 105 & 751 & 3,602 & 4,889 \\
    jedis & 749 & 6,219 & 27,404 & 12,132 \\
    jenkins-2.287 & 2,432 & 14,775 & 120,382 & 24,269 \\
    jitwatch-1.4.2 & 538 & 7,346 & 46,527 & 3,185 \\
    jsoup-1.14.2 & 246 & 1,551 & 17,449 & 11,235 \\
    junit4-4.13.2 & 310 & 1,541 & 10,769 & 8,533 \\
    libgdx-gdx-1.9.14 & 2,714 & 39,338 & 208,028 & 1,263 \\
    mall-1.0.2 & 747 & 14,322 & 100,990 & 81,223 \\
    mybatis-3.5.6 & 378 & 2,582 & 20,533 & 20,179 \\
    nanohttpd-2.3.1 & 75 & 405 & 3,821 & 7,129 \\
    netty-socketio-1.7.18 & 138 & 712 & 5,217 & 6,980 \\
    redisson-3.15.3 & 1,613 & 13,607 & 82,104 & 23,946 \\
    retrofit-1.6.0 & 118 & 403 & 4,790 & 43,653 \\
    rocketmq-4.9.2 & 996 & 7,536 & 70,871 & 21,990 \\
    Sa-Token-1.28.0 & 191 & 1,600 & 8,848 & 18,043 \\
    Sentinel-1.8.3 & 1,029 & 4,963 & 41,366 & 22,852 \\
    spring-cloud-alibaba-2.2.2 & 411 & 2,003 & 13,594 & 28,662 \\
    webmagic-develop-0.7.6 & 207 & 955 & 6,757 & 11,600 \\
    xxl-job-2.3.0 & 150 & 741 & 8,374 & 29,124 \\
    zxing-3.4.1 & 303 & 1,783 & 23,614 & 33,495 \\
    \midrule
    \textbf{Total} & 50,774 & 385,046 & 3,508,637 & 682,312 \\
    \bottomrule \\
  \end{tabular}
\end{table}

Table~\ref{tab:detection-tools} lists the four static analysis tools used to detect the nine different types of code smells~\citep{Santana2024}: \textit{Data Class} (DC), \textit{Dispersed Coupling} (DiCo), \textit{Feature Envy} (FE), \textit{Intensive Coupling} (IC), \textit{Long Method} (LM), \textit{Large Class} (LC), \textit{Long Parameter List} (LPL), \textit{Refused Bequest} (RB) and \textit{Shotgun Surgery} (SS). A \say{\faCheck} mark in the table represents the tool of choice for detecting a specific smell. For instance, we used JDeodorant and JSpIRIT to detect \textit{Large Class}. Individual outputs of the four static analysis tools used in our previous study are also available~\citep{Santana2024}. We use them in this paper to support comparisons between LLMs and static analysis tools. We chose these tools because they have shown good accuracy in detecting code smells in previous studies~\citep{fernandes2016review, fokaefs2011jdeodorant, madeyski2023detecting, paiva2017evaluation, santos2023yet}. 

\begin{table}[ht]
  \centering
  \caption{Static Analysis Tools Used to Detect Code Smells}
  \label{tab:detection-tools}
  \small
  \setlength{\tabcolsep}{4pt}
  \begin{tabularx}{\columnwidth}{@{} l *{9}{>{\centering\arraybackslash}X} @{}}
    \toprule
    \textbf{Tool}
      & \textbf{DC}
      & \textbf{DiCo}
      & \textbf{FE}
      & \textbf{IC}
      & \textbf{LC}
      & \textbf{LM}
      & \textbf{LPL}
      & \textbf{RB}
      & \textbf{SS} \\
    \midrule
    JDeodorant
      &            
      &            
      & \faCheck  
      &            
      & \faCheck  
      & \faCheck  
      &            
      &            
      &            
      \\
    PMD       
      & \faCheck  
      &            
      &            
      &            
      &            
      &            
      & \faCheck  
      &            
      &            
      \\
    Organic   
      & \faCheck  
      & \faCheck  
      & \faCheck  
      & \faCheck  
      &            
      & \faCheck  
      & \faCheck  
      & \faCheck  
      & \faCheck  
      \\
    JSpIRIT   
      &            
      & \faCheck  
      &            
      & \faCheck  
      & \faCheck  
      &            
      &            
      & \faCheck  
      & \faCheck  
      \\
    \bottomrule \\
  \end{tabularx}
\end{table}

\section{An Extended Dataset of Code Smells}
\label{sec:dataset}

\begin{figure*}[ht]
    \centering
    \includegraphics[width=\textwidth]{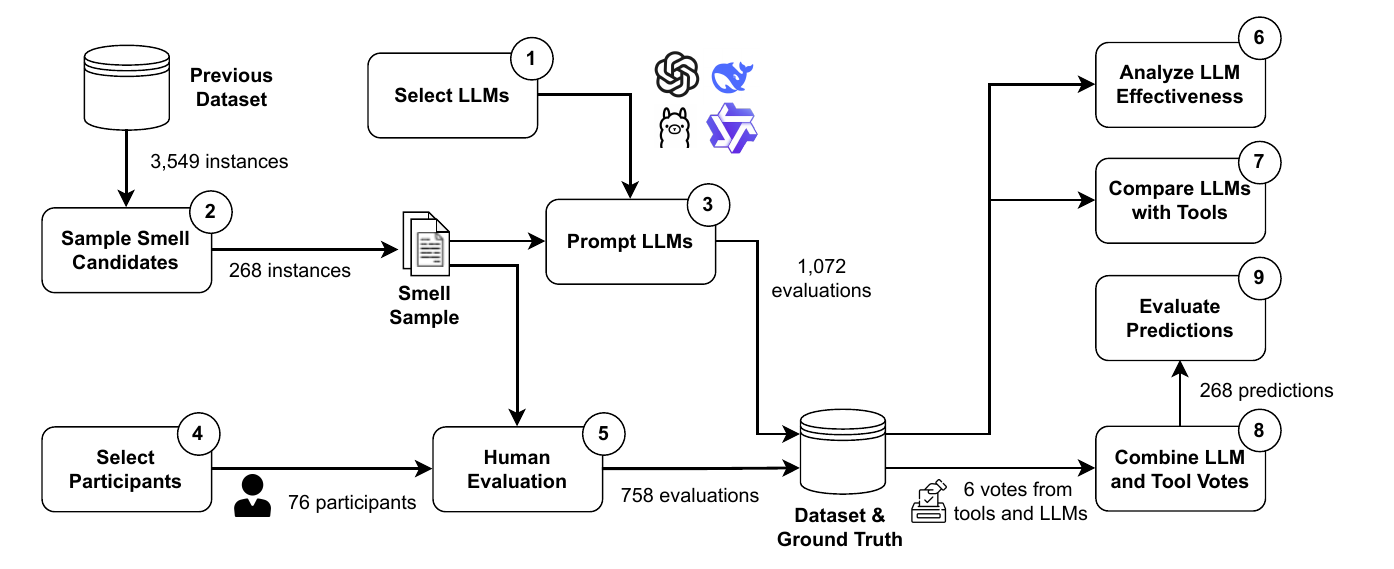}
    \caption{Steps of our study.}
    \label{fig:ultimate_dataset_v2}
\end{figure*}

Figure~\ref{fig:ultimate_dataset_v2} presents the steps we followed in this paper.
Steps 1 to 5 outline how we extended the previous dataset~\citep{Santana2024} by incorporating LLM-generated code-smell information and human evaluations for this study. The following sections describe these steps.

\subsection{Selected Large Language Models}
\label{sec:study-design:subsec:prompt}

The first step of our study (Step 1 in Figure~\ref{fig:ultimate_dataset_v2}) is to select the LLMs~\citep{baltes2025guidelines}. Table~\ref{tab:llms-infos} shows important information about the LLMs selected in this study: OpenAI's GPT-5 mini, Meta's Llama-3.3-70B-Instruct, DeepSeek's DeepSeek-R1-Distill-Qwen-32B, and Qwen's Qwen2.5-Coder-32B-Instruct. All models have similar features regarding context windows and knowledge cutoffs. For instance, DeepSeek-R1 and Qwen2.5-Coder both feature 32.5 billion parameters and a large context window of 131,000 tokens. Their knowledge cutoffs are July 2024 and March 2024, respectively. Llama-3.3 has 70 billion parameters and a context window of 128,000 tokens. Although GPT-5 mini does not reveal its parameter count, it has a similar context window of 128,000 tokens.

\begin{table}[ht]
\centering
\caption{Selected Large Language Models}
\label{tab:llms-infos}
    \begin{tabularx}{\linewidth}{l c c c}
        \toprule
        \textbf{Model} & \textbf{Parameters} & \textbf{Context Window} & \textbf{Knowledge Cutoff} \\
        \midrule
            DeepSeek-R1 & 32.5B & 131K & July 2024 \\
            GPT-5 mini & N/A & 128K & May 2024 \\
            Llama-3.3 & 70B & 128K & Dec 2023 \\
            Qwen2.5-Coder & 32.5B & 131K & Mar 2024 \\
        \bottomrule
    \end{tabularx}
\end{table}

We chose these LLMs because they are widely used in recent research~\citep{Caumartin2025,dong2025chatgpt,hou2024large,tanzil2024chatgpt,xue2024does}. For instance, GPT models have attracted a lot of attention from researchers, with studies covering many topics of software engineering, such as code generation~\citep{liu2024no}, computer science education~\citep{xue2024does}, refactoring~\citep{dong2025chatgpt}, code review~\citep{guo2024exploring}, test case generation~\citep{yuan2024evaluating}, and library selection~\citep{tanzil2024chatgpt}. We also included Llama-3.3 and DeepSeek-R1 because they are open source, allowing us to compare them with closed-source models, such as GPT-5 mini. For instance, in May of 2026, developers downloaded DeepSeek-R1 over 4 million times on HuggingFace, while Llama-3.3 reached more than 690,000 downloads, demonstrating strong interest in these models from the developer community. In addition, we included the code-generation model Qwen2.5-Coder to provide a different perspective on our study. Qwen2.5-Coder was the highest-ranked code model on Hugging Face's leaderboard in May of 2026~\footnote{https://huggingface.co/spaces/bigcode/bigcode-models-leaderboard}.

Given the knowledge cut-off dates of the selected models, which range from December 2023 to July 2024, we considered the possibility of data contamination, since some of the code analyzed in this study may have been available during model training~\citep{baltes2025guidelines,xia2024automated}. We mitigated this risk by relying on fixed snapshots of 30 active, top-starred Java projects from GitHub, all of which had recent commits and represented real-world codebases rather than synthetic samples~\citep{dozono2025should, rodriguez2025snipgen}. In addition, we rely on the same dataset to ensure that no particular bias was introduced for any model~\citep{huang2025comprehensive}. This approach does not eliminate contamination entirely, but reduces its impact and preserves the fairness of the comparison~\citep{baltes2025guidelines, fan2023automated,xia2023automated}.

\subsection{Sampling Code Smells} 
\label{subsec:sample-collection}

In this step, we randomly sampled 268 instances of code smells as candidates for this study (Step 2 in Figure~\ref{fig:ultimate_dataset_v2}). We defined the sample size based on three factors: balance, representativeness, and the viability of human validation. For representativeness and balance, we selected the same number of smells for each system and smell type in our empirical study. That is, we selected one instance of each smell in each system; 9 * 30 = 270, but we removed two duplicates. The duplicates have the same code in the methods of different systems. The sample size was determined by constraints on enrolling a large group of participants to manually evaluate the code smell instances. 

The selected sample includes both smelly and non-smelly classes and methods~\citep{Santana2024}. These groups are balanced to allow both static analysis tools and LLMs to distinguish smelly from non-smelly instances. Although six types of code smells in this study are found at the method level, we decided to provide their surrounding class as additional context to the LLMs.. This decision gives the models more context to help them detect code smells. Such context information could be particularly important for some smells that rely on inheritance relationships (e.g., \textit{Refused Bequest}) and class dependencies (e.g., \textit{Shotgun Surgery}). 

\subsection{The Used Prompt}
\label{sec:study-design:subsec:prompt-new}

\begin{figure*}[ht]
    \centering
    \includegraphics[height=0.55\textheight]{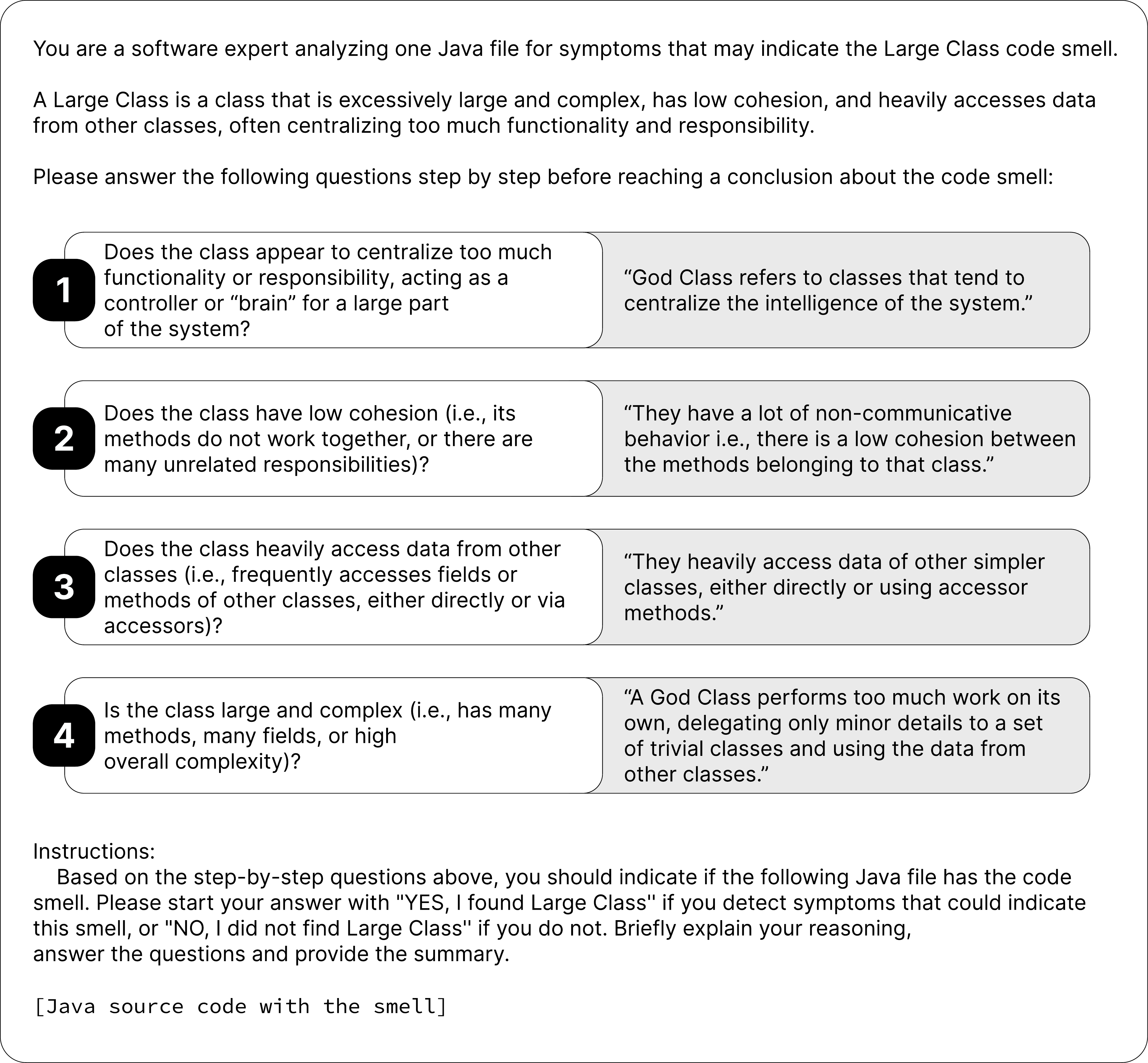}
    \caption{Prompt used to detect Large Class}
    \label{box:prompt-LC}
\end{figure*}

Step 3 in Figure~\ref{fig:ultimate_dataset_v2} is prompting the four LLMs to detect code smells. Figure~\ref{box:prompt-LC} illustrates the structured prompt template used in this step, with \textit{Large Class} shown as a representative example. We rely on Chain-of-Thought (CoT) prompting, which involves a series of intermediate reasoning steps~\citep{wei2022chain}. This prompting strategy aims to improve a model's capacity to produce organized, insightful answers. To guide the models' reasoning, we formulated four questions for each code smell in our study, using the definitions and detection strategies provided by~\citet{lanza2007object}. This book contains dedicated sections for all evaluated smells, except \textit{Long Parameter List}. In this case, we relied on the detection strategies defined by \citet{bhave2022deep}. This step ensures that the detection strategies in our prompts align with the established literature-based criteria for each code smell in our study.

The gray boxes in Figure~\ref{box:prompt-LC} present the quotes from the~\citet{lanza2007object} book for the Large Class code smell, while the box on the left shows the derived questions we used in our prompt. More precisely, Section 5.3 of their book~\citep{lanza2007object} proposes a detection strategy for this smell by identifying three key symptoms: (1)~\say{the class centralizes system intelligence, or its functional complexity is very high}, (2)~\say{the class cohesion is low}, and (3)~\say{the class uses many attributes from other classes}. Each symptom is accompanied by detailed explanations summarized in the first three items of Figure~\ref{box:prompt-LC}. In addition, we provide a general description of the smell, which contributes to a more comprehensive understanding. This general description motivated the inclusion of an additional, broader question, which is illustrated by item (4) in Figure~\ref{box:prompt-LC}. The prompts we used for the other eight smells are available in our replication package~\citep{baltes2025guidelines, dataset}. We systematically applied this process to both derive detailed detection criteria and define the questions used across all studied code smells. 


Because GPT-5 mini does not support custom temperature settings~\footnote{https://platform.openai.com/docs/guides/latest-model}, we used its default temperature of 1.0. To ensure consistency and facilitate fair comparison across all models, we applied this same temperature setting to all four models in our study. Other empirical studies in software engineering used similar temperature values~\citep{sagodi2025program, zhang2025can, ye2025prompt}. Finally, we limited all LLMs' outputs to 1,500 characters to simplify our analysis.

\subsection{Ground Truth Creation} 
\label{subsec:datasetsteps}

The goal of our ground truth is to represent the human perspectives of the analyzed code smells. As shown in Step 4 of Figure~\ref{fig:ultimate_dataset_v2}, we first selected 76 participants to manually evaluate the 268 code smell candidates. These participants are undergraduate Computer Science students in their final year. A background assessment confirmed that they had sufficient experience with Java programming. This assessment asked participants to indicate which Software Engineering topics they were familiar with from a predefined list, describe their professional experience, and self-assess their proficiency in both Java programming and code smell detection. Skill levels were rated on a five-point likert scale, from (1)~\say{Never heard of it} to (5)~\say{I’m an expert}. A complete breakdown of the participants' demographics and the forms used in the empirical experiment is available in our replication package~\citep{dataset}. We then offered participants a preparatory 1-hour lecture to ensure they had a clear understanding of code smells before participating in the evaluation~\citep{baltes2025guidelines}. After the controlled experiment, we created nine individual ground truths of manually validated classes, one for each code smell selected for this study. Each ground truth is represented by a distinct subset of 29 or 30 instances, depending on the code smell.

In Step 5, for each code smell, we asked participants to rate a class code on a five-point scale from 1 (it is definitely not a smell) to 5 (it is definitely a smell). Each participant was asked to evaluate only 10 randomly sampled classes for different code smells, yet two participants evaluated nine classes each. It is important to note that at least two participants evaluated each class. We consider a true positive in the ground truth if the average evaluator score exceeds 3 out of 5. Otherwise, the class is considered non-smelly. To ensure valid comparisons, all LLMs, human annotators, and static analysis tools detect the same code smells instances (see Section~\ref{subsec:sample-collection}). 

In general, the participants reported having good knowledge of software engineering concepts but were still relatively early in their industrial experience. The most familiar software engineering topic was Object-Oriented Programming, with 93.4\% of participants reporting familiarity, followed by Databases, Web Technologies, and Refactoring. LLMs in Software Development were also mentioned, but by a smaller group of participants. Regarding professional experience, 46.1\% of participants had at least 1 year in software development, and most reported Java proficiency (68.4\%). A similar pattern emerged for code smell detection, with the vast majority of participants (97.4\%) indicating at least some familiarity with the concept.
\section{Research Method}
\label{sec:research-method}

This section details the research questions of our study. It also describes the evaluation steps taken to assess LLMs for code smell detection and details the data analysis methods used to interpret our results.

\subsection{Goal and Research Questions} 
\label{subsec:rqs}

This study aims to evaluate the effectiveness of LLMs in detecting code smells in software projects. We also analyzed whether a mixed strategy combining LLMs and static analysis tools improves the code smell detection. To achieve these goals, we defined the following three research questions (RQs).

\begin{itemize}
  \item \textbf{RQ1:} How effective are LLMs in detecting code smells? 
  \item \textbf{RQ2:} How do LLMs compare with static analysis tools in code smell detection?
  \item \textbf{RQ3:} How effective is automated code smell detection when combining outputs from LLMs and tools?
\end{itemize}

\subsection{Evaluation Steps}
\label{subsec:studysteps}

Steps 6 to 9 in Figure~\ref{fig:ultimate_dataset_v2} depict the analyses we performed to answer the research questions in this study. In these steps, we rely on the dataset and corresponding ground truth (see Section~\ref{subsec:datasetsteps}) to evaluate the effectiveness of LLMs for code smell detection using three metrics: Recall, Precision, and F1-Score~\citep{goutte2005probabilistic}. To answer RQ1, Step 6 directly compares the LLM outputs with the ground truth created from 758 human evaluations.

In Step 7 of Figure~\ref{fig:ultimate_dataset_v2}, we answer RQ2 by comparing the effectiveness of LLMs with the static analysis tools used as baselines. The effectiveness of each LLM and tool is measured by its alignment with the human perspective on the analyzed code smells. Finally, we use Recall, Precision, and F1-Score to determine whether a combined strategy presented in Section~\ref{subsec:prediction} is an effective solution for code smell detection.

\subsection{Combining Outputs of Tools and LLMs} 
\label{subsec:prediction}

We propose a combined strategy in Step 8 of Figure~\ref{fig:ultimate_dataset_v2}, using a voting mechanism that takes as input the scores from the tools and LLMs used in this study. More precisely, this voting strategy combines the outputs of each code smell from two static analysis tools and four LLMs (6 votes)~\citep{mining2006data}. We should note that, although we rely on the outputs of four static analysis tools, only two of them were used for each smell, as detailed in Section~\ref{sec:background:original-dataset}. A code smell was predicted if it received at least three votes for that smell. We chose this number after empirically evaluating the results with different values and observing a higher correlation between the ground truth and the combined predictions. All votes carry equal weight, ensuring that both tools and LLMs have the same influence on the final decision regarding a smell candidate.

Table~\ref{tab:ground-truth} summarizes the sample of code smell candidates from the used dataset (``Sample''), the identified true positives in the ground truth (``Ground Truth''), and the detected ones by the proposed strategy (``Combined Strategy''). Our results reveal an imbalance in the number of code smells, consistent with previous studies~\citep{fernandes2016review,fokaefs2011jdeodorant,madeyski2020mlcq}. Notably, combined predictions generally flag more code smells than human evaluators. For instance, in six out of the nine smells, the number of instances detected in the combined predictions exceeds the manually annotated ground truth. However, easily detectable code smells, such as \textit{Long Method} and \textit{Large Class}, are commonly present in both the ground truth and the combined results. In contrast, more complex smells, such as \textit{Refused Bequest} and \textit{Shotgun Surgery}, are less represented. This distribution highlights the persistent challenges of reliably detecting certain types of code smells, whether evaluated by humans or automated strategies.

\begin{table}[ht]
\centering
\caption{Code smells in the Datasets}
\label{tab:ground-truth}
    \begin{tabularx}{\linewidth}{
        >{\hsize=1.3\hsize\raggedright\arraybackslash}X 
        >{\hsize=0.4\hsize\centering\arraybackslash}X   
        >{\hsize=1.1\hsize\centering\arraybackslash}X   
        >{\hsize=1.2\hsize\centering\arraybackslash}X   
    }
    \toprule
        \textbf{Code Smell} & \textbf{Sample} & \textbf{Ground Truth} &
        \textbf{Combined Strategy} \\
        \midrule
            Data Class          & 29 & 17 & 21 \\
            Dispersed Coupling  & 30 & 18 & 22 \\
            Feature Envy        & 29 & 16 & 24 \\
            Intensive Coupling  & 30 & 19 & 25 \\
            Large Class         & 30 & 20 & 23 \\
            Long Method         & 30 & 25 & 22 \\
            Long Parameter List & 30 & 16 & 25 \\
            Refused Bequest     & 30 & 10 & 10 \\
            Shotgun Surgery     & 30 & 16 & 16 \\
        \bottomrule
    \end{tabularx}
\end{table}

\section{Results}
\label{sec:results}

This section presents the main findings of this paper, focusing on the most interesting results. The section is structured according to our three research questions.

\subsection{Effectiveness of LLMs in Code Smell Detection (RQ1)}
\label{subsec:toolcomparison}

\begin{table}[ht]
  \centering
  \caption{Effectiveness Metrics per Model and Tool}
  \label{tab:eff-all-smells-humans}
  \footnotesize 
  \setlength{\tabcolsep}{2pt} 
  
  \begin{tabular}{l l cccc cccc}
  \toprule
  & & \multicolumn{4}{c}{\textbf{LLMs}} & \multicolumn{4}{c}{\textbf{Tools}} \\
  \cmidrule(lr){3-6} \cmidrule(lr){7-10}
  \textbf{Smell} & \textbf{Metric} & \textbf{DS-R1} & \textbf{GPT-5} & \textbf{Lla-3.3} & \textbf{Qwen} & \textbf{JDeo} & \textbf{JSpI} & \textbf{Org} & \textbf{PMD} \\
  \midrule
  \multirow{3}{2cm}{Data Class}
    & Precision & 0.88 & 0.82 & \textbf{0.94} & \textbf{0.94} & - & - & 0.71 & 0.82 \\
    & Recall  & 0.83 & \textbf{0.88} & 0.84 & 0.80 & - & - & 0.71 & 0.74 \\
    & F1-score    & 0.86 & 0.85 & \textbf{0.89} & 0.86 & - & - & 0.71 & 0.78 \\
  \midrule
  \multirow{3}{2cm}{Dispersed Coupling}
    & Precision & 0.33 & 0.33 & 0.78 & 0.78 & - & \textbf{0.89} & 0.61 & - \\
    & Recall  & 0.75 & 0.60 & 0.67 & 0.67 & - & 0.62 & \textbf{0.79} & - \\
    & F1-score    & 0.46 & 0.43 & 0.72 & 0.72 & - & \textbf{0.73} & 0.69 & - \\
  \midrule
  \multirow{3}{2cm}{Feature Envy}
    & Precision & 0.50 & 0.06 & \textbf{0.94} & 0.50 & 0.81 & - & 0.75 & - \\
    & Recall  & 0.50 & 0.50 & \textbf{0.58} & 0.42 & 0.50 & - & 0.55 & - \\
    & F1-score    & 0.50 & 0.11 & \textbf{0.71} & 0.46 & 0.62 & - & 0.63 & - \\
  \midrule
  \multirow{3}{2cm}{Intensive Coupling}
    & Precision & 0.79 & 0.53 & 0.63 & \textbf{0.84} & - & 0.74 & 0.63 & - \\
    & Recall  & \textbf{0.83} & 0.71 & 0.63 & 0.73 & - & 0.61 & 0.67 & - \\
    & F1-score    & \textbf{0.81} & 0.61 & 0.63 & 0.78 & - & 0.67 & 0.65 & - \\
  \midrule
  \multirow{3}{2cm}{Large Class}
    & Precision & 0.80 & 0.50 & \textbf{0.95} & 0.85 & 0.70 & 0.60 & - & - \\
    & Recall  & 0.80 & 0.83 & 0.83 & 0.85 & 0.61 & \textbf{0.92} & - & - \\
    & F1-score    & 0.80 & 0.63 & \textbf{0.88} & 0.85 & 0.65 & 0.63 & - & - \\
  \midrule
  \multirow{3}{2cm}{Long Method}
    & Precision & 0.52 & \textbf{0.80} & \textbf{0.80} & 0.52 & 0.76 & - & 0.72 & - \\
    & Recall  & 0.93 & 0.95 & 0.80 & 0.93 & 0.83 & - & \textbf{1.00} & - \\
    & F1-score    & 0.67 & \textbf{0.87} & 0.80 & 0.67 & 0.79 & - & 0.84 & - \\
  \midrule
  \multirow{3}{2cm}{Long Parameter List}
    & Precision & 0.56 & 0.63 & 0.88 & 0.50 & - & - & \textbf{1.00} & 0.25 \\
    & Recall  & \textbf{0.69} & 0.59 & 0.54 & 0.47 & - & - & 0.57 & 0.67 \\
    & F1-score    & 0.62 & 0.61 & 0.67 & 0.48 & - & - & \textbf{0.73} & 0.36 \\
  \midrule
  \multirow{3}{2cm}{Refused Bequest}
    & Precision & 0.40 & 0.10 & 0.20 & 0.30 & - & \textbf{0.50} & \textbf{0.50} & - \\
    & Recall  & 0.36 & 0.33 & 0.18 & 0.30 & - & \textbf{0.42} & 0.29 & - \\
    & F1-score    & 0.38 & 0.15 & 0.19 & 0.30 & - & \textbf{0.45} & 0.37 & - \\
  \midrule
  \multirow{3}{2cm}{Shotgun Surgery}
    & Precision & 0.56 & 0.69 & 0.50 & 0.00 & - & \textbf{0.75} & 0.44 & - \\
    & Recall  & \textbf{0.69} & 0.58 & 0.67 & 0.00 & - & 0.63 & 0.50 & - \\
    & F1-score    & 0.62 & 0.63 & 0.57 & 0.00 & - & \textbf{0.69} & 0.47 & - \\
  \bottomrule
  \end{tabular}
\end{table}

Table~\ref{tab:eff-all-smells-humans} (first 6 columns) presents the effectiveness in terms of Precision, Recall, and F1-Score of each LLM---DeepSeek-R1, GPT-5 mini, Llama-3.3, and Qwen2.5-Coder---across the code smells analyzed in this study. We highlight in \textbf{bold} the best automated strategy, either an LLM or a tool, for the respective metric and smell. Overall, our results indicate that LLMs achieve the strongest and most consistent performance across all metrics for structurally simpler smells, namely \textit{Data Class, Large Class, and Long Method}. For instance, except for GPT-5 mini, which has a Precision of 0.50, the other three LLMs excel at identifying \textit{Large Class} with high Precision: 0.95 for Llama-3.3, 0.85 for Qwen2.5-Coder, and 0.80 for DeepSeek-R1. Similarly, all models achieved a Recall above 0.80 for these three smells, successfully retrieving a broad set of relevant instances. With respect to F1-scores, all LLMs are also highly effective with values above 0.8, with just a few exceptions. 

For a second group of three code smells, \textit{Dispersed Coupling, Feature Envy, and Intensive Coupling}, some LLMs present good results while others fail. For instance, Llama-3.3 and Qwen2.5-Coder achieve high Precision (0.78 each) and F1-Score (0.72 each) for \textit{Dispersed Coupling}, while DeepSeek-R1 and GPT-5 mini show substantial drops in effectiveness (e.g., F1 below 0.5), suggesting that these models are not well-suited for detecting this code smell. We can draw similar observations for the other two smells in this group, although the best results are achieved with different LLMs. That is, Llama-3.3 yields better results for \textit{Feature Envy}, and DeepSeek-R1 yields better results for \textit{Intensive Coupling}. In summary, LLM performance is more variable for these three code smells. Therefore, we should carefully select the appropriate LLM that fits our purpose. 

When it comes to the most subjective and context-dependent code smells, such as \textit{Long Parameter List, Refused Bequest, and Shotgun Surgery}, all models face substantial challenges. Apart from Llama-3.3 with a Precision of 0.88 for \textit{Long Parameter List}, no model has achieved 0.70 or higher on any metric for these three smells. For \textit{Long Parameter List}, Llama-3.3 achieves relatively high Precision (0.88) but lower Recall (0.54), yielding an F1-score of 0.67.  DeepSeek-R1, on the other hand, has a Recall of 0.69, but a low Precision of 0.56. However, the poor effectiveness of LLMs is especially noticeable for \textit{Refused Bequest}, where Precision and Recall both decline, leading all models to score below 0.40 on F1-score. The inherent subjectivity and semantic complexity of these smells, where code context rather than static metrics guides detection, pose significant limitations for LLM-based detections. For instance, \textit{Shotgun Surgery} often involves code scattered across multiple modules, a nuance that requires deep, holistic reasoning that current LLMs rarely replicate.





\begin{framed}
\noindent
    \textbf{RQ1 Findings}: Overall, the results show that smells based on clear syntatic patterns, such as size, class structure, and strong coupling between classes, are easier for LLMs to detect. In contrast, smells that require a deeper understanding of the code's context are more challenging for these models. These findings imply that while LLMs can reliably identify certain code smells, there is still work to be done in helping them detect more subjective and complex issues in code.
\end{framed}

\subsection{Comparison of LLMs with Static Analysis Tools (RQ2)}
\label{subsec:efficiency}

The four rightmost columns of Table \ref{tab:eff-all-smells-humans} present the results obtained from static analysis tools. Together with the previous columns reporting LLM results, we can perform a direct comparison between static analysis tools and LLM-based approaches.
Our results show that LLMs often outperform static analysis tools for at least three code smells: \textit{Data Class, Feature Envy, and Intensive Coupling}.
For \textit{Data Class}, all LLMs are more effective than static analysis tools for all three metrics, indicating that LLMs are clearly the best choice for this smell.   
In the case of \textit{Feature Envy}, Llama-3.3 fares best with a notably higher Precision (0.94) and F1-score (0.71). However, other LLMs perform poorly on this smell, revealing inconsistencies across models.
For \textit{Intensive Coupling}, JSpIRIT achieves an F1-score of 0.67, which is higher than two LLMs (GPT-5 and Llama-3.3). However, DeepSeek-R1 and Qwen2.5-Coder are better options, suggesting that these LLMs are effective at identifying code smells related to intensive use of method calls. That is, DeepSeek-R1 outperforms both JSpIRIT and Organic in terms of F1-score (0.81) and Recall (0.83), while Qwen2.5-Coder achieves the highest Precision (0.84).

For well-defined code smells, such as \textit{Large Class and Long Method}, both LLMs and static analysis tools often achieve strong results. In fact, the best-performing LLMs tend to outperform or closely match the tools for these smells. For instance, Llama-3.3 achieves the highest Precision (0.95) and F1-score (0.88) for \textit{Large Class}, surpassing JDeodorant and JSpIRIT. However, JSpIRIT outperforms all other strategies in Recall (0.92). 
Results for \textit{Long Method} follow a similar pattern, but GPT-5 mini shows the highest Precision (0.80) and F1-score (0.87), while Organic achieves perfect Recall (1.00). These results confirm the effectiveness of LLMs in detecting size-related code smells in par with traditional tools.

In contrast, code smells lacking a clear structure or with strong context dependence---e.g.,~\textit{Dispersed Coupling, Long Parameter List, Refused Bequest, and Shotgun Surgery}---present higher variability in detection results, while static tools keep a consistent advantage. For instance,
JSpIRIT achieves high Precision (0.89) and an F1-score of 0.73 for \textit{Dispersed Coupling}, outperforming all LLMs. Similarly, the Organic tool is the best strategy for \textit{Long Parameter List}, with high Precision (1.00) and F1-score (0.73). On the other hand, \textit{Refused Bequest} and \textit{Shotgun Surgery} remain particularly challenging for both LLMs and static analysis tools. JSpIRIT achieves the highest F1-scores for these two smells (0.45 and 0.69, respectively), far higher than all LLMs. This result shows that while LLMs are expanding the boundaries of automated code analysis, specialized static analysis tools still provide more effective detection for nuanced or scattered patterns that require deeper code context or semantic reasoning.




\begin{framed}
\noindent
    \textbf{RQ2 Findings}: The results show that LLMs tend to fare better than static analysis tools in detecting simpler and size-based smells. While they are more effective than these tools for five out of the nine smells in this study, traditional tools remain competitive for more subtle or semantically complex code smells. This finding highlights the complementary nature of both strategies and the potential to leverage LLM strengths alongside the established benefits of static analysis tools.
\end{framed}

\subsection{A Combined Prediction Strategy for Code Smells (RQ3)}
\label{subsec:humancomparison}

Table~\ref{tab:best-vs-automated} directly compares, for each code smell, the effectiveness of the best individual strategy with the combined strategy proposed in Section~\ref{subsec:prediction}. The LLM or static analysis tool with the highest F1-Score (Table~\ref{tab:eff-all-smells-humans}) was selected as the best strategy. This comparison aims to answer RQ3, that is, to assess the effectiveness of a detection strategy that combines outputs from LLMs and tools. It may also provide valuable insights into which strategy a researcher or practitioner might prefer, depending on their specific detection priorities. 

\begin{table}[ht]
  \centering
  \caption{Comparison of the Best Strategy and Combined Prediction}
  \label{tab:best-vs-automated}
  \begin{tabularx}{\linewidth}{l*{6}{>{\centering\arraybackslash}X}}
    \toprule
    & \multicolumn{3}{c}{\textbf{Best Strategy}} 
    & \multicolumn{3}{c}{\textbf{Combined Prediction}} \\
    \cmidrule(lr){2-4} \cmidrule(lr){5-7}
    \textbf{Code Smell} & \textbf{P} & \textbf{R} & \textbf{F1}
    & \textbf{P} & \textbf{R} & \textbf{F1} \\
    \midrule
    Data Class & \textbf{0.94} & 0.84 & \textbf{0.89} & 0.81 & \textbf{1.00} & \textbf{0.89} \\
    Dispersed Coupling & \textbf{0.89} & 0.79 & 0.62 & 0.73 & \textbf{0.89} & \textbf{0.80} \\
    Feature Envy & \textbf{0.94} & 0.58 & \textbf{0.71} & 0.50 & \textbf{0.75} & 0.60 \\
    Intensive Coupling & \textbf{0.79} & 0.83 & \textbf{0.81} & 0.68 & \textbf{0.89} & 0.77 \\
    Large Class & \textbf{0.95} & 0.83 & 0.88 & 0.87 & \textbf{1.00} & \textbf{0.93} \\
    Long Method & 0.80 & \textbf{0.95} & 0.87 & \textbf{0.95} & 0.84 & \textbf{0.89} \\
    Long Parameter List & \textbf{1.00} & 0.57 & \textbf{0.73} & 0.56 & \textbf{0.88} & 0.68 \\
    Refused Bequest & \textbf{0.50} & \textbf{0.42} & \textbf{0.45} & 0.30 & 0.30 & 0.30 \\
    Shotgun Surgery & \textbf{0.75} & 0.63 & \textbf{0.69} & 0.69 & \textbf{0.69} & \textbf{0.69} \\
    \bottomrule
  \end{tabularx}
\end{table}

For three code smells, namely \textit{Dispersed Coupling, Large Class, and Long Method}, the combination of multiple tools and LLMs produces the best F1-scores. A common characteristic of these smells is that they are related to localized structures and size anti-patterns. For instance, both \textit{Large Class} and \textit{Long Method} achieve their highest F1-scores with the combined prediction (0.93 and 0.89, respectively), compared to their best individual strategies (0.88 and 0.87, respectively). In these cases, the combined strategy delivers the best balance between Recall and Precision, detecting almost all true instances (1.00 for \textit{Large Class} and 0.84 for \textit{Long Method}) while still maintaining high Precision (0.87 for \textit{Large Class} and 0.95 for \textit{Long Method}). For these simpler, localized smells, the combined prediction strategy is the most effective detection method, making it an attractive choice for researchers and practitioners seeking comprehensive, robust results. 

Two code smells---\textit{Data Class} and \textit{Shotgun Surgery}---do not show a clear advantage of the combined prediction strategy over the best individual strategy. Both approaches yield equivalent F1-Scores (0.89 for \textit{Data Class} and 0.69 for \textit{Shotgun Surgery}), indicating that these strategies produce consistent results. 
However, it is important to note that the researcher or practitioner must know in advance which individual-detection strategy is best. Therefore, a combined strategy may be a more robust option for these two code smells.

Conversely, four code smells---\textit{Feature Envy, Intensive Coupling, Long Parameter List, and Refused Bequest}---remain better detected by a single top-performing tool or LLM. Interestingly, LLMs are the most effective for \textit{Feature Envy} (Llama-3.3) and Intensive Coupling (DeepSeek-R1) while traditional tools fare best for \textit{Long Parameter List} (Organic) and \textit{Refused Bequest} (JSpIRIT). For instance, \textit{Feature Envy} achieves much higher F1-score (0.71) and Precision (0.94) with Llama-3.3 than with the combined strategy (0.60 and 0.50). On the other hand, \textit{Long Parameter List} is best detected by Organic, as it achieves perfect Precision (1.00) and the highest F1-score (0.73 versus 0.68 with the combined strategy). \textit{Intensive Coupling} and \textit{Refused Bequest} also achieved higher F1-scores with individual strategies (0.81 vs 0.77, and 0.45 vs 0.30, respectively). For these four code smells, sticking with the most effective single strategy provides more targeted and effective detection.

Finally, it is important to note that the combined prediction strategy achieves the best Recall across 7 of 9 code smells. Therefore, it seems a rational choice for researchers or practitioners seeking many true instances at the expense of lower Precision. This result is especially valuable for those who prioritize automatic smell detection followed by manual validation. However, it can also lead to a higher rate of false positives and thus higher validation cost for some complex or subjective smells.

\begin{framed}
\noindent
    \textbf{RQ3 Findings}: A direct comparison with the best individual strategy shows that the combined prediction performs best for code smells with localized structures, such as \textit{Large Class} and \textit{Long Method}, achieving the highest F1 and Recall scores. For more subjective or context-dependent smells, such as \textit{Feature Envy} and \textit{Refused Bequest}, the best individual strategy yields better results. However, when the optimal strategy is unknown, the combined strategy offers a robust alternative. Finally, the optimal strategy depends on whether Recall or Precision is the main priority. In the former case, LLMs are a better option for seven code smells. 
\end{framed}
\section{Discussion}
\label{sec:discussion}

This section discusses the main findings of this study, the possible uses of the dataset created, and the implications for software practitioners and researchers.

\subsection{Main Findings} 
\label{subsec:summary-results}

\newcommand{\happyicon}{%
  \includegraphics[height=1.05em]{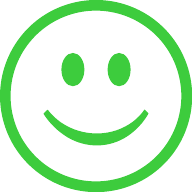}%
}

\newcommand{\neutralicon}{%
  \includegraphics[height=1.05em]{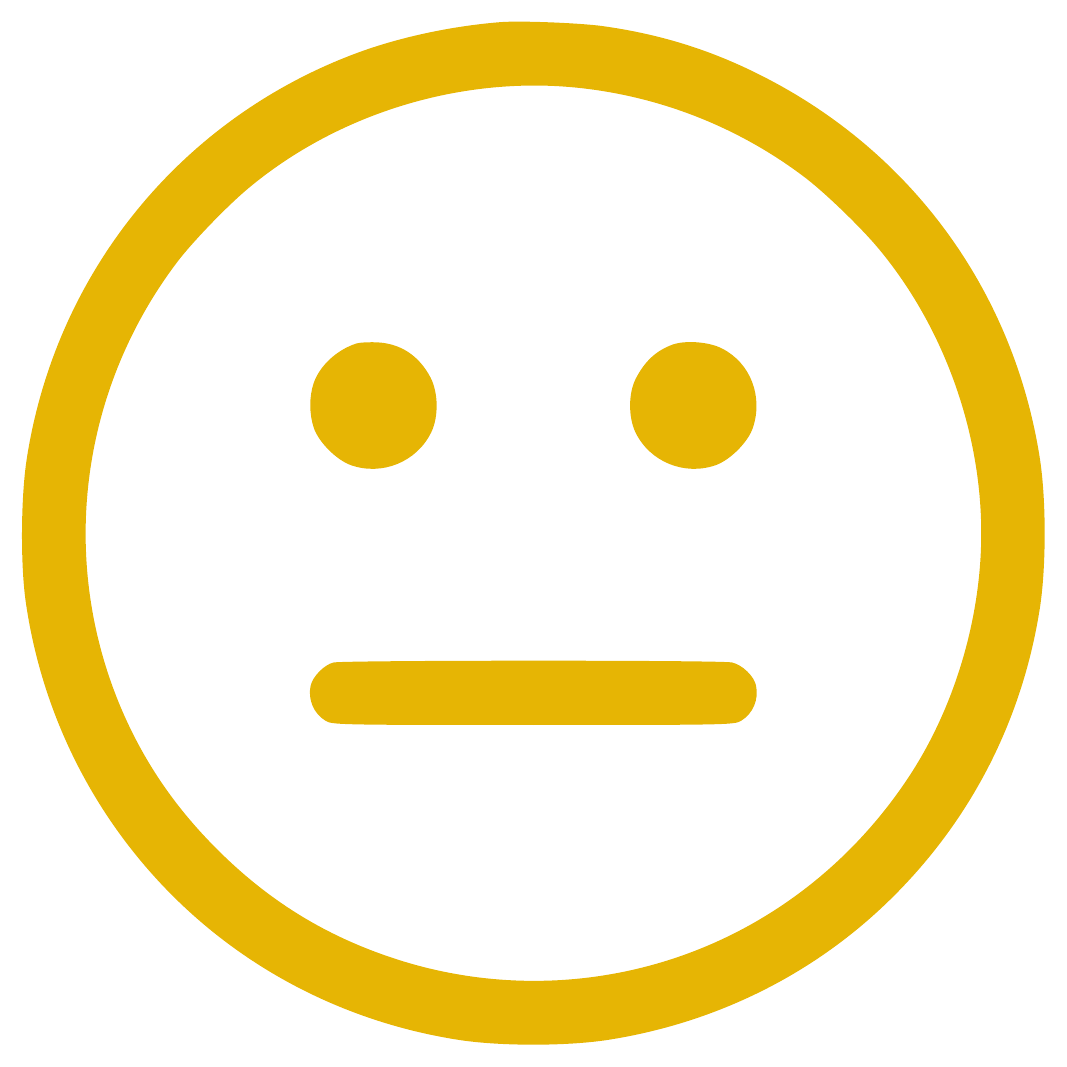}%
}

\newcommand{\sadicon}{%
  \includegraphics[height=1.05em]{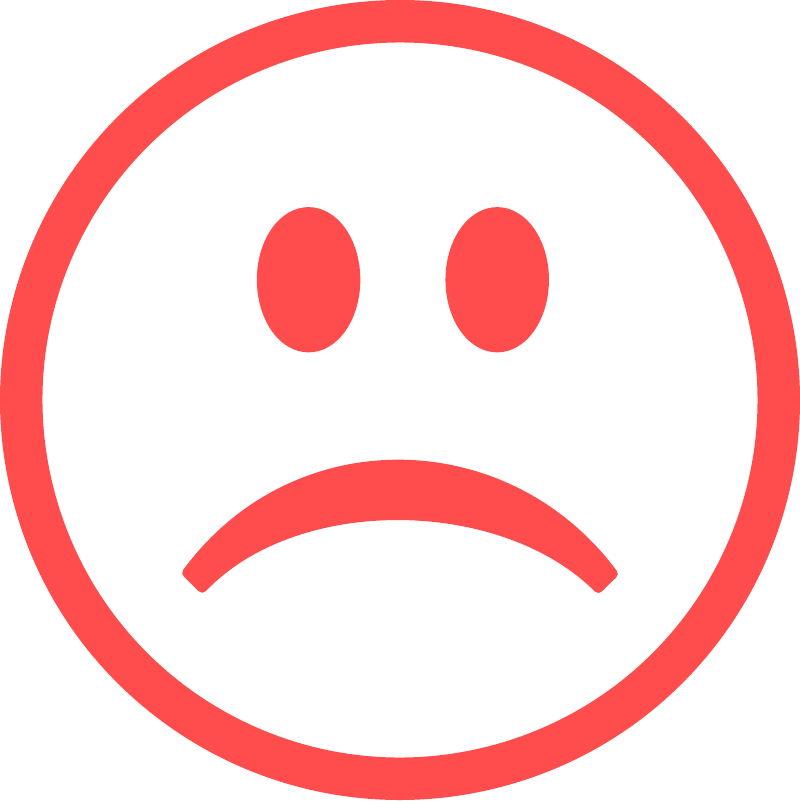}%
}

\newcommand{\medalicon}{%
  \includegraphics[height=1.05em]{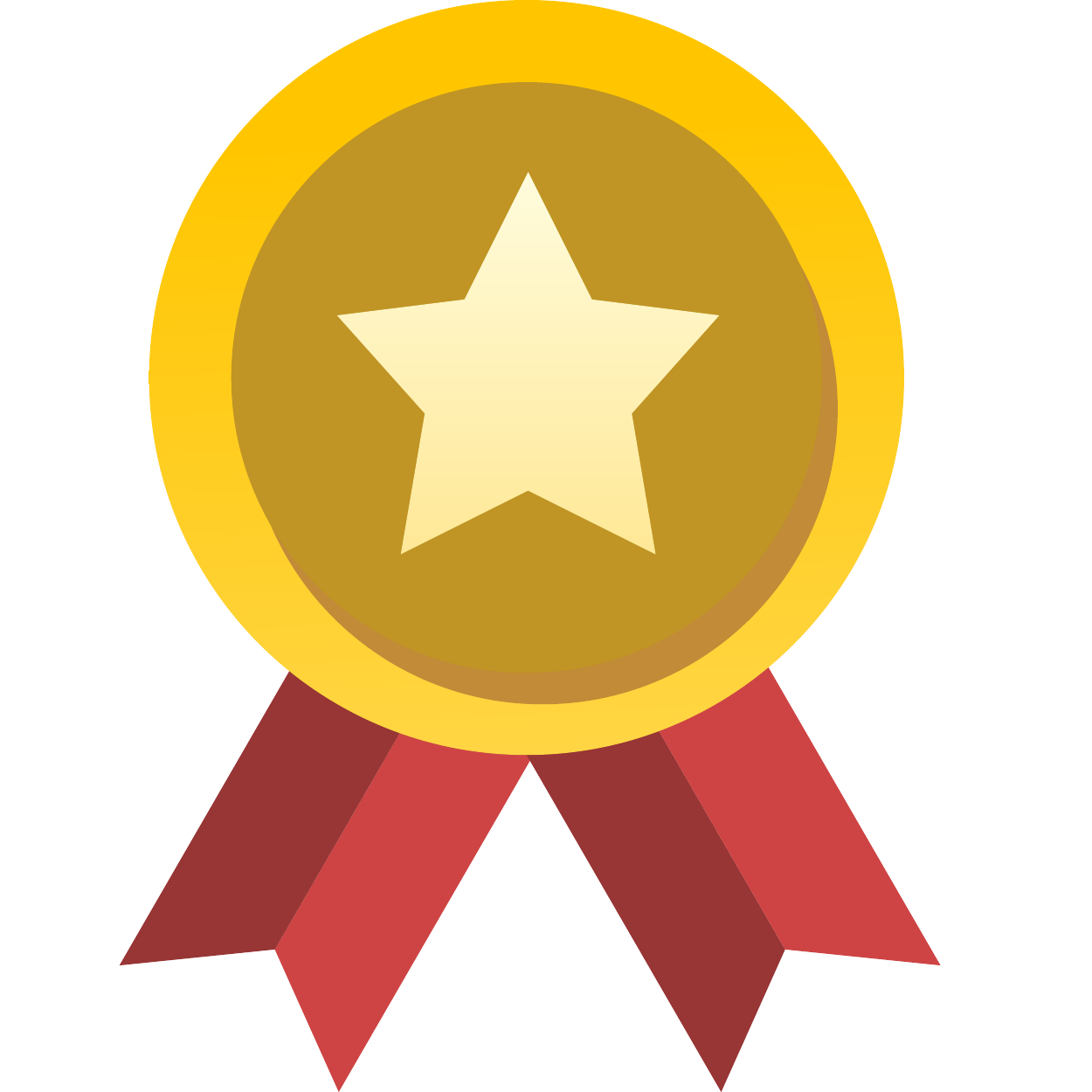}%
}

\newcommand{\bestperformer}[1]{\medalicon~#1}

\begin{table*}[t]
  \centering
  \caption{Effectiveness Metrics for all strategies with Top Results}
  \label{tab:eff-all-smells-summary}

  \footnotesize
  \setlength{\tabcolsep}{2.5pt}
  \renewcommand{\arraystretch}{0.98}

  \begin{adjustbox}{max width=\textwidth}
    \begin{tabular}{@{} l *{9}{c} @{}}
      \toprule
      & \multicolumn{4}{c}{\textbf{Large Language Models}}
      & \multicolumn{4}{c}{\textbf{Static Analysis Tools}}
      & \makecell{\textbf{Combined}\\\textbf{Predictions}} \\
      \cmidrule(lr){2-5}
      \cmidrule(lr){6-9}
      \textbf{Smell}
        & \textbf{DeepSeek}
        & \textbf{GPT}
        & \textbf{Llama}
        & \textbf{Qwen}
        & \textbf{JDeodorant}
        & \textbf{JSpIRIT}
        & \textbf{Organic}
        & \textbf{PMD}
        & \multicolumn{1}{c}{} \\
      \midrule
      Data Class
        & \neutralicon
        & \neutralicon
        & \bestperformer{\happyicon}
        & \neutralicon
        &
        &
        & \neutralicon
        & \neutralicon
        & \bestperformer{\happyicon}
        \\
      \addlinespace[1pt]
      Dispersed Coupling
        & \sadicon
        & \sadicon
        & \neutralicon
        & \neutralicon
        &
        & \neutralicon
        & \neutralicon
        &
        & \bestperformer{\happyicon}
        \\
      \addlinespace[1pt]
      Feature Envy
        & \sadicon
        & \sadicon
        & \bestperformer{\neutralicon}
        & \sadicon
        & \neutralicon
        &
        & \neutralicon
        &
        & \neutralicon
        \\
      \addlinespace[1pt]
      Intensive Coupling
        & \bestperformer{\happyicon}
        & \neutralicon
        & \neutralicon
        & \neutralicon
        &
        & \neutralicon
        & \neutralicon
        &
        & \neutralicon
        \\
      \addlinespace[1pt]
      Large Class
        & \happyicon
        & \neutralicon
        & \bestperformer{\happyicon}
        & \happyicon
        & \neutralicon
        & \neutralicon
        &
        &
        & \bestperformer{\happyicon}
        \\
      \addlinespace[1pt]
      Long Method
        & \neutralicon
        & \bestperformer{\happyicon}
        & \happyicon
        & \neutralicon
        & \happyicon
        &
        & \happyicon
        &
        & \bestperformer{\happyicon}
        \\
      \addlinespace[1pt]
      Long Parameter List
        & \neutralicon
        & \neutralicon
        & \neutralicon
        & \sadicon
        &
        &
        & \bestperformer{\neutralicon}
        & \sadicon
        & \neutralicon
        \\
      \addlinespace[1pt]
      Refused Bequest
        & \sadicon
        & \sadicon
        & \sadicon
        & \sadicon
        &
        & \bestperformer{\sadicon}
        & \sadicon
        &
        & \sadicon
        \\
      \addlinespace[1pt]
      Shotgun Surgery
        & \neutralicon
        & \neutralicon
        & \neutralicon
        & \sadicon
        &
        & \bestperformer{\neutralicon}
        & \neutralicon
        &
        & \bestperformer{\neutralicon}
        \\
      \bottomrule
    \end{tabular}
  \end{adjustbox}

  \vspace{0.75ex}
  {\footnotesize
    \textbf{Legend:}
    \raisebox{-0.4em}{\happyicon\ }~(F1 $\geq$ 0.80)\quad
    \raisebox{-0.4em}{\neutralicon\ }~(0.51 $\leq$ F1 $<$ 0.80)\quad
    \raisebox{-0.4em}{\sadicon\ }~(F1 $\leq$ 0.50)\quad
    \raisebox{-0.4em}{\medalicon\ } Best strategies
  }
\end{table*}

Table~\ref{tab:eff-all-smells-summary} presents a summary of the effectiveness of each static analysis tool, LLM, and the combined prediction for all code smells analyzed in this study. We rely on the F1-score for each strategy and smell to directly compare their effectiveness. To help interpret the results, we use emojis and a color scale: green indicates high performance (F1-score between 0.80 and 1.00), yellow denotes moderate effectiveness (0.51–0.79), and red signals limited performance (0.50 or below). A check mark indicates the best F1-score for each smell, with ties indicated when multiple strategies share the top F1-score. This table can serve as a practical guide for researchers and practitioners looking to select the most effective detection strategy for each specific code smell.

Several other findings emerged from our study. For instance, one finding relates to code smells with localized structural characteristics, such as \textit{Dispersed Coupling}, \textit{Large Class}, and \textit{Long Method}. For these smells, the combined prediction strategy based on majority voting achieved the highest F1 Scores and Recall compared to both the best single LLMs and tools. For instance, both Large Class and Long Method achieved near-perfect Recall and high Precision when leveraging a combined approach. These results show that, for smells directly related to size or clear structural patterns, integrating multiple detection sources helps minimize missed cases and produces robust, highly effective results.

Our evaluation did not identify a clear advantage for either detection strategy for \textit{Data Class} and \textit{Shotgun Surgery}. Both the combined prediction and the best individual strategy (Llama-3.3 and JSpIRIT, respectively) yielded equivalent F1 Scores. This tied effectiveness means developers and researchers can select either strategy based on their workflow or preferences. However, the combined strategy does not require determining the best individual one in advance and thus offers a practical alternative. 

Our findings are more nuanced for smells such as \textit{Feature Envy}, \textit{Intensive Coupling}, \textit{Long Parameter List}, and \textit{Refused Bequest}. For these smells, a specialized tool or an LLM often produced the best results, rather than the combined-prediction strategy. For instance, Llama-3.3 performed best at detecting Feature Envy, achieving much higher Precision and F1-score than the combined predictions. Similarly, Organic outperformed all other strategies for Long Parameter List, achieving perfect Precision. These results suggest that, for more subjective or semantically complex smells, specialization and explicit knowledge of inter-class coupling, responsibility allocation, and inheritance hierarchies still outperform the combined prediction strategy, which can sometimes lead to conflicting predictions and reduce overall accuracy.

Figure~\ref{fig:long-parameter-list-example} shows an instance of \textit{Long Parameter List} included in the ground truth of this study. The two method declarations contain several parameters, which made the smell apparent to the static analysis tools and to the participants. However, none of the individual LLMs identified this instance. One possible explanation is that the snippet resembles a concise API-oriented declaration, which may have reduced the smell's visibility when considered in isolation. Another possibility is that the models relied more on local code semantics than on the explicit parameter count used by traditional detection criteria. In addition, without a broader class or project context, the LLMs may have interpreted the declarations as acceptable within a framework-style implementation.

\begin{figure}[ht]
    \centering
    \begin{lstlisting}[style=codeStyle]
  package com.alibaba.fastjson.asm;

  public interface MethodVisitor {
    ...

    void visitFieldInsn(int opcode, String owner, String name, String desc);
    
    void visitMethodInsn(int opcode, String owner, String name, String desc);

    ...
  } \end{lstlisting}
    \caption{Example of Long Parameter List fragments that were flagged as smelly}
    \label{fig:long-parameter-list-example}
\end{figure}

\begin{figure}[ht]
    \centering
    \begin{lstlisting}[style=codeStyle]
  package hudson.console;

  import hudson.util.ByteArrayOutputStream2;

  import java.io.IOException;
  import java.io.OutputStream;

  public abstract class LineTransformationOutputStream extends OutputStream {

    ...

    public abstract static class Delegating extends LineTransformationOutputStream {

      protected final OutputStream out;

      protected Delegating(OutputStream out) {
        this.out = out;
      }

      @Override
      public void flush() throws IOException {
        out.flush();
      }

      @Override
      public void close() throws IOException {
        super.close();
        out.close();
      }
    }
} \end{lstlisting}
    \caption{Example of Refused Bequest fragment that was flagged as smelly}
    \label{fig:refused-bequest-example}
\end{figure}

Figure~\ref{fig:refused-bequest-example} shows an example of Refused Bequest from our ground truth. The~\textit{LineTransformationOutputStream} class extends~\textit{OutputStream}, but its behavior is highly specialized around line buffering and end-of-line processing, which may make the inheritance relationship appear weak or unnatural. This instance was flagged by the static analysis tools and by the participants, but none of the individual LLMs identified it. One possible explanation is that the class still implements the expected stream operations, which may have made inheritance acceptable to the models despite the design mismatch. Another possibility is that this smell instance is more closely related to structural flaws than to semantic misunderstanding, although recognizing Refused Bequest often requires understanding whether the subclass truly benefits from the inherited interface and behavior. In this case, the models may have focused on the concrete method implementations rather than on the broader inheritance fit.

When comparing LLMs and static analysis tools, our results indicate that LLMs have matured enough to match or exceed traditional tools for six out of nine smells, especially those grounded in clear metrics or modular class designs, such as \textit{Data Class}, \textit{Feature Envy}, and \textit{Intensive Coupling}. However, they continue to struggle where deeper contextual or semantic understanding is necessary. This finding is more evident for \textit{Long Parameter List}, \textit{Refused Bequest}, and \textit{Shotgun Surgery}, which remain challenging for both LLMs and static tools.

An interesting trend is the systematic improvement of Recall when multiple strategies are combined in a single detection strategy. The combined prediction strategy consistently increased the number of true positive detections, which can be highly beneficial in contexts where missing code smells are a greater risk than occasionally signaling false positives. This finding highlights the strength of the combined prediction strategy as a safer option, especially in automated workflows followed by human review.

Finally, we found that model effectiveness varied not only across code smells but also across LLMs. Llama-3.3 and DeepSeek-R1 often delivered the most effective results among LLMs, suggesting that differences in the architecture of the LLMs and training data can have a measurable impact on their effectiveness for code smell detection. This observation highlights the importance of model selection and the potential value in continuing to fine-tune or customize LLMs for code smell analysis.

\subsection{Expected Use of Our Dataset} 
\label{subsec:datasetimplications}

Our dataset, built with LLM-detected and human-validated code smells, offers several promising avenues for both researchers and practitioners. By providing not only the raw source code but also detailed detection outputs from multiple LLMs, static analysis tools, and human votes for each class, the dataset becomes a valuable resource for comparison, benchmarking, and the development of new detection strategies.

For researchers, the dataset offers a unique opportunity to investigate the effectiveness of different code smell detectors. As shown in our results, some smells, such as \textit{Large Class} and \textit{Long Method}, can be reliably detected using a combination of automated strategies, whereas others, such as \textit{Feature Envy} and \textit{Refused Bequest}, remain challenging to detect. The performance per-smell of the dataset can be used to motivate and evaluate improvements in smell detection models, the development of more refined LLM prompting strategies, agent architectures, or approaches that incorporate richer code semantics and project context. Researchers can also leverage the chain-of-thought rationale captured in the LLM prompts and outputs to create new detection models.

For practitioners and tool developers, the dataset offers a practical reference for verifying and tuning static analysis tools, LLM-integrated coding assistants, or IDE extensions. Developers working on large codebases can benefit from the dataset's examples of both successful and problematic detections, using them as benchmarks for their own tools or as test cases for CI/CD pipelines. The evaluation provided by human evaluators, established as ground truth, also provides a gold standard against which new tooling can be evaluated and calibrated.

Notably, the structure of our dataset allows for a straightforward extension. Future releases could expand the range of code smells covered, add more languages beyond Java, or include more recent or larger LLMs as they become available. With well-documented prompt templates and evaluation scales, researchers can easily add their own experiments, swap in new models, or repeat our process with code from different domains or repositories. Similarly, collecting additional human votes or involving more experienced or diverse software developers could enrich the reliability and depth of the ground truth.

Another prospective use for our dataset is in educational settings. Instructors can use the varied annotated examples to illustrate standard code smells, help students understand the nuances of different detection methods, or train students in both automated and manual code review practices. The inclusion of multiple perspectives in the dataset (humans, LLMs, and static analysis tools) makes it an excellent resource for critical thinking and for teaching best practices in software engineering.

\subsection{Implications for Researchers}
\label{subsec:researchers}

Our findings open several avenues for future research in AI-driven code smell detection. One of the most interesting research opportunities lies in bridging the performance gap for subjective and context-dependent code smells. Despite the increasing capabilities of LLMs and combined detection in structurally explicit smells, such as \textit{Large Class} and \textit{Long Method}, complex cases, such as \textit{Feature Envy} and \textit{Refused Bequest}, remain persistently challenging. Researchers could focus on enriching model context, investigating hybrid models that blend code analysis with design documentation, or building smarter prompting strategies that push LLMs to better understand intent, rationale, and domain-specific patterns beyond structural metrics.

Furthermore, our results reveal the inherent limitations of relying solely on combined predictions for nuanced smells. They also raise questions about the best way to integrate or weight conflicting signals from heterogeneous detectors in a combined or ensemble strategy. Future research might explore adaptive weighting strategies or estimation frameworks that leverage not only the final predictions, but also the reasoning steps provided as~\say{chain-of-thought} explanations. Comparing the robustness and transparency of such hybrid approaches with classical voting could be an important next step.

Another promising direction is advancing explainable AI in code analysis. Our dataset provides detailed~\say{chain-of-thought} reasoning tied to each LLM evaluation. Therefore, researchers can mine these rationales, analyze patterns in LLM logic, and correlate the explanations generated by the models with the correctness of code-smell detections. Insights from the LLM responses could help create a path towards models that not only flag code quality issues but also justify their suggestions in a way developers can trust.

Finally, our work encourages the exploration of transfer learning and cross-domain evaluation. Since our detection pipeline and prompt methodology are open and fully described, researchers could adapt and apply them to new languages and frameworks. By comparing how models perform across different codebases or programming languages, researchers can better understand the extent of AI models' code knowledge and identify areas where adaptation to specific domains is needed.

\subsection{Implications for Practitioners}
\label{subsec:practitioners}

Our findings provide clear, actionable guidance for practitioners seeking to integrate automated code smell detection into their development processes. One of the most notable practical takeaways is how the effectiveness of a detection strategy depends on the specific type of smell being targeted. For code smells with well-defined, easily quantifiable structures, such as \textit{Large Class} and \textit{Long Method}, practitioners can confidently rely on a prediction strategy that combines several LLMs and tools. These strategies deliver consistently high Recall and F1-score, ensuring that even rare or subtle instances are less likely to escape detection. In legacy systems or software where safety is crucial, high coverage could be prioritized. In these cases, combining detection strategies can help mitigate the risk of smelly code remaining undetected.

However, for more contextual or nuanced code smells, such as \textit{Dispersed Coupling}, \textit{Feature Envy}, and \textit{Long Parameter List}, our results stress the importance of appropriate tool selection. That is, our study shows that a single, specialized tool or well-chosen LLM, such as Llama-3.3 for Feature Envy or Organic for Long Parameter List, offers more accurate, consistent, and interpretable results than collective voting. Selecting the right tool for such smells helps developers avoid \say{alert fatigue} from false positives and can streamline improvement efforts, saving teams both time and cognitive load.

Practitioners should also weigh the trade-off between missing true smells (false negatives) and generating excessive warnings (false positives). Our results indicate that combined predictions, while strong in Recall, can generate more candidate issues, potentially increasing the cost of manual review. For teams with limited bandwidth or where the developer's trust in automated tooling is paramount, focusing on high-Precision single-method detection for more complex smells is likely to deliver more value. However, projects prioritizing early warning and quality gates may gain greater assurance from combined predictions, even if some follow-up evaluation is required.

Finally, the usability of our dataset extends beyond running benchmarks, enabling practitioners to experiment, calibrate, and continuously improve their chosen strategies. By referencing detailed records of where and why each strategy succeeded or failed, development teams can tune static analysis rules, customize LLM prompt designs, or even blend tool and LLM predictions in custom-made workflows. The traceability and transparency of our results support informed decision-making and promote a learning loop for continuous quality improvement.

\subsection{Practical Implications of LLM-Augmented Code Smell Detection}
\label{subsec:pratical-implications}

In most real-world settings, teams already rely on at least one static analysis tool as part of code smell detection~\citep{tsantalis2008jdeodorant}, code review~\citep{ahmed2017senticr}, continuous integration/deployment~\citep{ghaleb2026ci,ichtsis2022merging}, or periodic quality checks~\citep{rostami2023usage,zhang2023survey}. Our findings suggest that LLMs can be incorporated as a complementary stage in this workflow, especially to revisit the smell candidates produced by these tools and to enrich the analysis with a broader contextual understanding of the code. This view is also consistent with other software engineering tasks, where the combination of LLMs and traditional tools has shown promise, such as test generation and software vulnerability detection~\citep{munson2025little,szabo2023incrementalizing,yang2025advancing, yang2025knighter}. In this scenario, LLMs do not replace existing detectors but rather help extend their coverage, particularly for smells where combining different detection signals improves the identification of relevant instances, such as Dispersed Coupling and Shotgun Surgery.

Another practical implication concerns the explanatory potential of LLMs. Unlike a simple warning produced by a static rule, the reasoning generated by an LLM can help developers understand why a smell is present and which design characteristics make the code problematic. This added explanation can be useful not only to support the immediate review of a smell candidate, but also to help developers learn how to reduce the recurrence of similar issues in future changes. In this sense, the model can serve as a support mechanism for both diagnosis and prevention, making code-smell reports more informative and easier to translate into refactoring actions.

Finally, the adoption of LLMs should be guided by cost considerations. Running an LLM over an entire codebase may be unnecessary in many projects, especially when a static analysis tool is already available to narrow down the set of suspicious instances~\citep{huang2025comprehensive}. A more cost-effective strategy is to apply the LLM selectively, for instance, to files touched by recent changes, to high-risk modules, or to candidates already flagged by a static analyzer. The combined prediction strategy can further support this goal by serving as a prioritization layer, since instances supported by multiple detectors can be reviewed first. In this way, the use of combined outputs can improve coverage while keeping human review effort and operational cost at manageable levels.
\section{Threats to Validity}
\label{sec:threats}

This section discusses the main validity concerns in our study, including internal, external, and construct validity, as well as reliability. We describe potential threats in each category and outline the steps taken to mitigate them.

\textbf{Internal validity} refers to factors that could affect the results of our study without our knowledge~\citep{wohlin2012}. An important threat is the sensitivity of LLM prompts and configurations, as small changes can yield different results. This feature can make it difficult to attribute the outcomes solely to the model's capabilities. To mitigate these threats, we used the same prompt design and temperature across all LLMs, reported all configuration details, and included all prompts in the replication package to ensure the findings are robust and reproducible.

Another threat to internal validity arises from the eventual bias introduced by the code samples used during LLM training, potentially affecting their ability to detect code smells in previously unseen code.
We mitigated this threat by using a dataset containing source code extracted from the 30 most starred GitHub repositories.
By selecting widely known code samples, we ensure that all models are equally likely to have been exposed to the same code during training. This approach promotes consistency in the evaluation.

\textbf{External validity} relates to the extent to which our findings can be generalized beyond the specific context of our study~\citep{wohlin2012}. Generalization to other languages and contexts is a threat to external validity, as results based solely on Java projects may not apply to other programming languages or software domains. Although we understand the limitations of our results, we made it possible to adapt the prompts used in this study to other languages and models since all prompts and scripts are available in our replication package~\citep{dataset}.

Another potential threat to the external validity of this study is the profile of the participants, as most were undergraduate students with limited professional experience and only a fair familiarity with code smell detection. To mitigate this threat, we first selected participants from the final year of the computer science curriculum who had already been exposed to key software engineering topics. We also provided a preparatory 1-hour lecture to ensure they shared a common understanding of code smells and the specific smell types analyzed in this study~\citep{baltes2025guidelines}.

\textbf{Construct validity} threats arise from how our results are established~\citep{wohlin2012}, such as the use of a voting system for the automated prediction strategy and the selection of efficiency metrics. These decisions can introduce bias into the observations. To mitigate them, we grounded our methodology in established practices and recommendations from peer-reviewed studies, ensuring alignment with widely accepted approaches in the field~\citep{aljamaan2021voting,goutte2005probabilistic,perez2020systematic,reis2022crowdsmelling}.

\textbf{Reliability} refers to how consistently our findings can be reproduced. If another group of researchers wants to repeat this study under similar conditions, the results need to be comparable. However, variability in responses from models such as GPT-5 mini, Llama-3.3, Qwen2.5-Coder, and DeepSeek-R1, as well as the subjective nature of human detection, could affect reliability. To minimize these threats, we standardized our evaluation procedures, applied the same dataset to all models, and thoroughly documented our methodology. These decisions allow future researchers to replicate our process and achieve similar outcomes. Furthermore, the reproducibility of our results depends on the exact versions of the models and the dataset used. To support reproducibility, we specified the versions of all models used in this study and made the dataset used in our experiments available, thereby providing a clear reference point for future studies.
\section{Related Work}
\label{sec:related-work}

Researchers are extensively exploring the use of LLMs to assist developers in several software engineering tasks, including code generation~\citep{Caumartin2025}, software migration~\citep{Ziftci2025, Wang2025, Almeida2024}, program repair~\citep{Kulsum2024}, code review~\citep{Sghaier2025}, and test case generation~\citep{alshahwan2024automated,Chen2024, schafer2023empirical}.
In this section, we provide an overview of studies on LLMs for software development and focus on their use for code smell detection.

\textbf{Static analysis tools for code smell detection.} Prior research has extensively studied code smell detection using static analysis tools~\citep{fernandes2016review,paiva2017evaluation,tsantalis2008jdeodorant}. \citet{fernandes2016review} conducted a systematic literature review and comparative study of code smell detection tools, revealing substantial overlap and redundancy among tools, as well as wide variation in recall and precision across smells. \citet{paiva2017evaluation} evaluated the accuracy and agreement of three popular tools, Deodorant, PMD, and JSpIRIT, showing that tool effectiveness strongly depends on the smell type and that high agreement often stems from identifying non-smelly entities rather than true positives. \citet{tsantalis2008jdeodorant} introduced JDeodorant, a refactoring-aware detection strategy that highlights the benefits of coupling code smell detection with concrete refactoring opportunities. Building upon these foundations, our study investigates whether modern LLMs can overcome some of the limitations observed in traditional tools and earlier machine learning models by leveraging richer contextual understanding of source code.

\textbf{LLMs have been largely used in software development.} Recent studies have explored the strengths and limitations of LLMs in code generation and related software engineering tasks~\citep{al2022readable,Caumartin2025,dong2024self,liu2024no,obrien2024prompt}. For example, \citet{liu2024no} systematically assessed ChatGPT’s code-generation capabilities, highlighting both its strengths and persistent vulnerabilities. \citet{dong2024self} proposed a self-collaboration framework in which multiple LLM agents, each with a specific software role, collaborate to improve code generation, achieving notable gains over single-agent approaches. \citet{Caumartin2025} showed that, with proper tuning, open-source Llama models can approach ChatGPT's performance in code refinement tasks. \citet{obrien2024prompt} found that prompt engineering with TODO comments can either help or hinder Copilot’s ability to address technical debt, depending on the clarity of the comments. \citet{al2022readable} found that Copilot’s generated code tends to be as readable as human code, but noted that programmers inspect it less, raising concerns about automation bias. Our work complements these efforts by focusing on detecting code smells with LLMs. 

\textbf{LLMs for code smell detection.} Some preliminary studies have also investigated how LLMs can be applied to detect and mitigate code smells in different settings~\citep{jiang2024unearthing,silva2024detecting,wu2024ismell}. For instance, \citet{silva2024detecting} examined ChatGPT's performance at identifying four classic code smells and found that explicitly naming the smells in the prompt improved detection accuracy, though difficulties persisted in more complex cases. \citet{wu2024ismell} introduced iSMELL, an ensemble technique that integrates LLMs with traditional code analysis tools, outperforming both individual LLMs and single expert systems in detecting and correcting specific code smells. \citet{jiang2024unearthing} investigated gas-wasting code smells in Ethereum smart contracts, employing GPT-4 to detect inefficiencies and recommend fixes that led to substantial cost reductions. In contrast, our study examines a broader range of Java code smells, compares proprietary and open-source LLMs, and provides a dataset that combines automated predictions and a human-validated ground truth.

\textbf{LLMs for code refactoring.} In addition to code smell detection, early research explored how LLMs can support code smell refactoring~\citep{choi2024iterative,nunes2025evaluating,pomian2024assist,shirafuji2023refactoring}. For instance, \citet{choi2024iterative} introduced an iterative approach in which ChatGPT 3.5 repeatedly refactors the most complex methods, leading to a steady drop in overall complexity. \citet{pomian2024assist} developed EM-Assist, an automated strategy that uses ChatGPT 3.5 to suggest and rank~\say{Extract Method} refactorings, achieving a 53\% recall for complex cases. \citet{shirafuji2023refactoring} presented a strategy to select optimal few-shot examples to guide ChatGPT 3.5 in reducing~\say{Cyclomatic Complexity}. More recently, \citet{nunes2025evaluating} empirically evaluated two LLMs, namely Copilot Chat and Llama 3.1, for their ability to automatically fix real-world maintainability issues in Java projects. Their results show that while LLMs can improve code readability and address a subset of code smells, they often introduce compilation errors or new maintainability issues. Unlike these studies, we do not focus on refactoring in this paper. However, our study can be seen as a previous step for improving code by first detecting (our focus) and then refactoring code smells.
\section{Conclusion and Future Work}
\label{sec:conclusion}

This study investigated how effective LLMs are at identifying a broad set of nine code smells in 30 Java software projects. In general, our results demonstrated that LLMs, when backed by careful prompt engineering, effectively detect code smells with clear structural patterns, such as those related to class size or complexity. Moreover, we also show that LLMs combined with traditional static analysis tools often match or surpass these strategies in isolation. However, for more subjective or context-dependent smells, such as Feature Envy and Refused Bequest, specialized tools or individual LLMs remain more effective. 

In future work, we plan to expand the set of analyzed LLMs by incorporating additional models, such as Sonnet, Claude, and Copilot, and to increase the number of code smells in our dataset. With a robust set of human-validated examples already available in our dataset, we highlight the potential for further research by using these high-quality instances to fine-tune LLMs and explore more advanced prompting techniques, such as few-shot learning. Ultimately, these improvements may yield a more robust assessment of LLMs' potential as reliable tools for automated code-smell detection in real-world software engineering contexts.

\section*{Declarations}

\subsection*{Funding:}

This research was partially supported by Brazilian funding agencies: CNPq (Grant 406089/2025-6), CAPES, and FAPEMIG (Grant APQ-01488-24).

\subsection*{Ethical approval:}

This item is not applicable.

\subsection*{Informed consent:}

All participants provided their written informed consent.

\subsection*{Author Contributions:}

All this paper’s experimental work was conducted by Saymon Souza. Amanda Santana, Eduardo Figueiredo, Igor Muzetti Pereira, João Eduardo Montandon and Lionel Briand contributed equally to the data analysis and paper writing.

\subsection*{Data Availability Statement:}

To repeat and/or refine and/or refute this work, see our scripts and data at ~\url{https://github.
com/sssouza/code_smell_detection_with_llms}

\subsection*{Conflict of Interest:}

The authors declared that they have no conflict of interest. The authors have no competing interests to declare that are relevant to the content of this article.

\subsection*{Clinical Trial Number:}

Not applicable.

\bibliographystyle{spbasic}
\bibliography{references}

\end{document}